\documentclass{moriond}

\usepackage{atlasphysics}
\usepackage{amsmath}

\def\Journal#1#2#3#4{{#1} {\bf #2}, #3 (#4)}

\def\PRD{{Phys. Rev.} D}

\def\JINST{JINST}
\def\EPJC{{Eur. Phys. J.} C}
\def\JHEP{JHEP}

\def\sst{\scriptscriptstyle}

\def\ra{\rightarrow}

\def\be{\begin{equation}}
\def\ee{\end{equation}}
\def\bea{\begin{eqnarray}}
\def\eea{\end{eqnarray}}

\newcommand{\Photo}{}

\begin{document}
%\linenumbers

\vspace*{4cm}
\title{Top Quark Mass Measurements at the LHC}

\author{T.G.McCarthy on behalf of the ATLAS and CMS Collaborations}

\address{Max-Planck-Institut f$\ddot{u}$r Physik, F$\ddot{o}$hringer Ring 6, 80805 M$\ddot{u}$nchen, Germany}

\maketitle\abstracts{
This document highlights a number of top-quark mass measurements carried out by the \\ ATLAS and CMS collaborations based on the combined LHC Run~1 datasets at centre-of-mass energies of $\sqrt{s}=7$  and $8$ \TeV.  A wide range of analysis strategies are employed for a number of final-state signatures.  Measurements of both the pole mass as well as the value of $m_{top}$ as defined by the Monte Carlo generator in simulated signal samples are discussed. 
}

\vspace{-0.8cm}
\section{Introduction}\label{sec:intro}

The top-quark mass ($m_{top}$) is a fundamental parameter of the standard model of particle physics (SM).  As the heaviest of all SM particles with a mass close to the electroweak symmetry-breaking scale, the top quark plays a pivotal role in the theory of elementary particles.  The exact value of the top-quark mass has implications on a number of theoretical predictions.  This motivates the need for precision measurements of $m_{top}$. 

The successful LHC data-taking at centre-of-mass energies of $\rts=$ 7 and 8 \TeV \ led to the production of tens of millions of top quarks \textit{via} proton-proton collisions at the centres of both the ATLAS \cite{atlas_paper} and CMS \cite{cms_paper} detectors.  The collection of such an unprecedented dataset ushered in an era of high-precision measurements of the top-quark mass at the LHC in which the sources of systematic uncertainty typically outweigh those statistical in nature.  The analyses presented in this document are divided into direct and indirect measurements.

\vspace{-0.2cm}
\section{Direct measurements from event reconstruction and kinematics}\label{sec:direct}

The direct measurements employ so-called ideogram \cite{cms_78tev_allchannels} or template methods, by which an observable sensitive to $m_{top}$ is constructed from various reconstructed objects.  In some cases it is possible to fully reconstruct one or two of the top quarks.  In other cases alternative observables are built from a number of possible kinematic variables.  For the standard template methods a set of templates, in the form of binned histograms, are then constructed for a number of discrete simulated Monte Carlo datasets which differ only in their input $m_{top}$ value.  These templates subsequently allow for an estimate of the top-quark mass to be made by means of a fit to the distribution of the same observable from the measured data.  The ideogram methods employed by CMS build separate parameterized shapes of the observables for the various jet-quark permutation cases in order to evaluate per-event likelihood values; the per-event likelihoods are combined and subsequently used to extract an estimate of the top-quark mass which is calibrated using simulated events.  The direct measurements are typically subdivided into various channels based on the final state decay products.

\vspace{-0.35cm}
\subsection{All-hadronic $t\bar{t}$ decay channel}\label{sec:allhad} 

Both CMS and ATLAS have performed measurements of $m_{top}$ specifically by identifying and reconstructing events in which a \ttbar \ pair is produced and decays to an all-hadronic final state.  The all-hadronic channel has the advantage of a large branching ratio and the absence of any immediate neutrinos from the \Wboson-boson decays.  Constributions from the dominant multi-jet background can be suppressed by applying tight event-selection cuts and dedicated tagging requirements to identify the presence of bottom-flavoured hadrons in reconstructed jets. In the $\rts=$7 \TeV \ ATLAS analysis \cite{atlas_7tev_allhad} the susceptibility of the $m_{top}$ measurement to uncertainties in the jet energy scale (JES) is suppressed by employing the ratio of the reconstructed top-quark to \Wboson-boson invariant mass (\rtt$=\mjjj/\mjj$) as the $m_{top}$-sensitive observable.

The analyses performed by CMS at both $\rts=$ 7 and 8 \TeV \ employ an ideogram method.\cite{cms_78tev_allchannels}  Comparisons in performance are made between a one- and two-dimensional fits.  In the latter case, a simultaneous fit is performed to distributions of both the reconstructed \Wboson-boson and top-quark mass.  This allows for a simultaneous measurement of $m_{top}$ and a global jet scale factor (JSF), which leads to a reduced systematic uncertainty due to the JES.

\vspace{-0.35cm}
\subsection{Dileptonic $t\bar{t}$ decay channel}\label{sec:dilep}

The dileptonic channel is characterized by very low backgrounds and a smaller branching ratio compared with those of the other standard decay channels.  A difficulty presents itself in terms of the ability to reconstruct the \ttbar \ system due to the presence of two non-interacting neutrinos in the final state.  A variety of analysis strategies are employed to overcome this complication.  The $\rts=$7 \TeV \ ATLAS measurement employs a one-dimensional template method, where the reconstructed invariant lepton-$b$-quark mass ($m_{\ell b}$) is chosen as the $m_{top}$ observable.\cite{atlas_7tev_semilep}

A number of measurements are performed in the dileptonic \ttbar \ channel by CMS.  A $\rts=$8 \TeV \ analysis similarly employs the $m_{\ell b}$ observable together with a forward folding of the theoretical prediction in order to perform a final fit to the data \cite{cms_8tev_dilep_mlb}.  A separate analysis strategy, chosen for both the $\rts =$7 and 8 \TeV \ datasets, makes use of an Analytical Matrix Weighting Technique (AMWT), whereby a weight is assigned to the various solutions of the kinematic equations in any given event.\cite{cms_78tev_allchannels} Yet another $\rts=$7 \TeV \ analysis extracts a measurement for $m_{top}$ based on the kinematic endpoints of certain key distributions.\cite{cms_7tev_kinendpoints} Finally, a measurement of $m_{top}$ is performed in a $\rts=$8 \TeV \ analysis based on the peak position of the $b$-jet energy spectrum alone.\cite{cms_8tev_dilep_bjetpeak}

\vspace{-0.35cm}
\subsection{Semileptonic $t\bar{t}$ decay channel}\label{sec:semilep}

\begin{figure}
\vspace{-0.55cm}
\begin{minipage}[c]{0.47\linewidth}
\centerline{\includegraphics[width=0.73\linewidth]{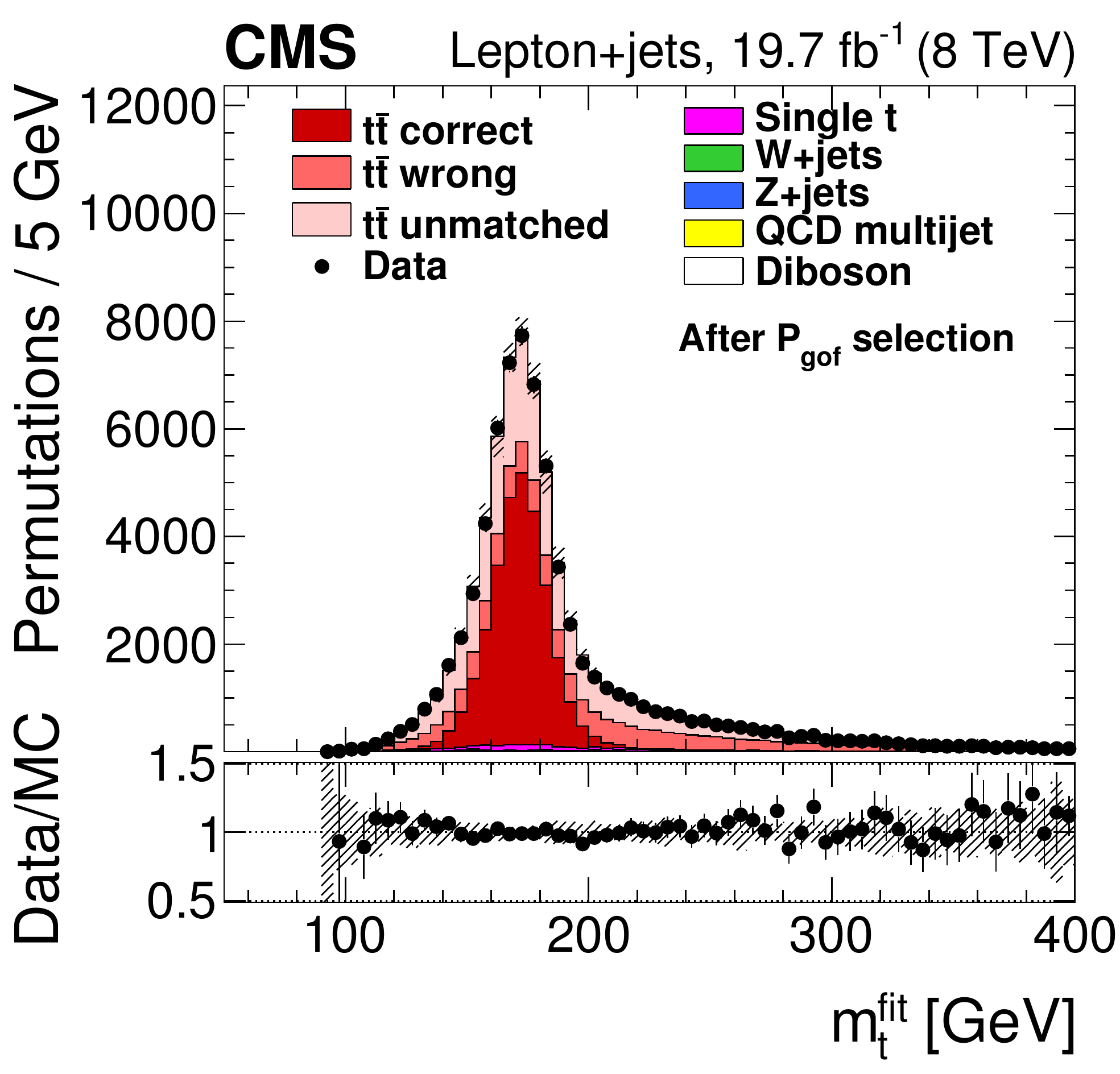}}
\end{minipage}
\hfill
\begin{minipage}[c]{0.47\linewidth}
\centerline{\includegraphics[width=0.88\linewidth]{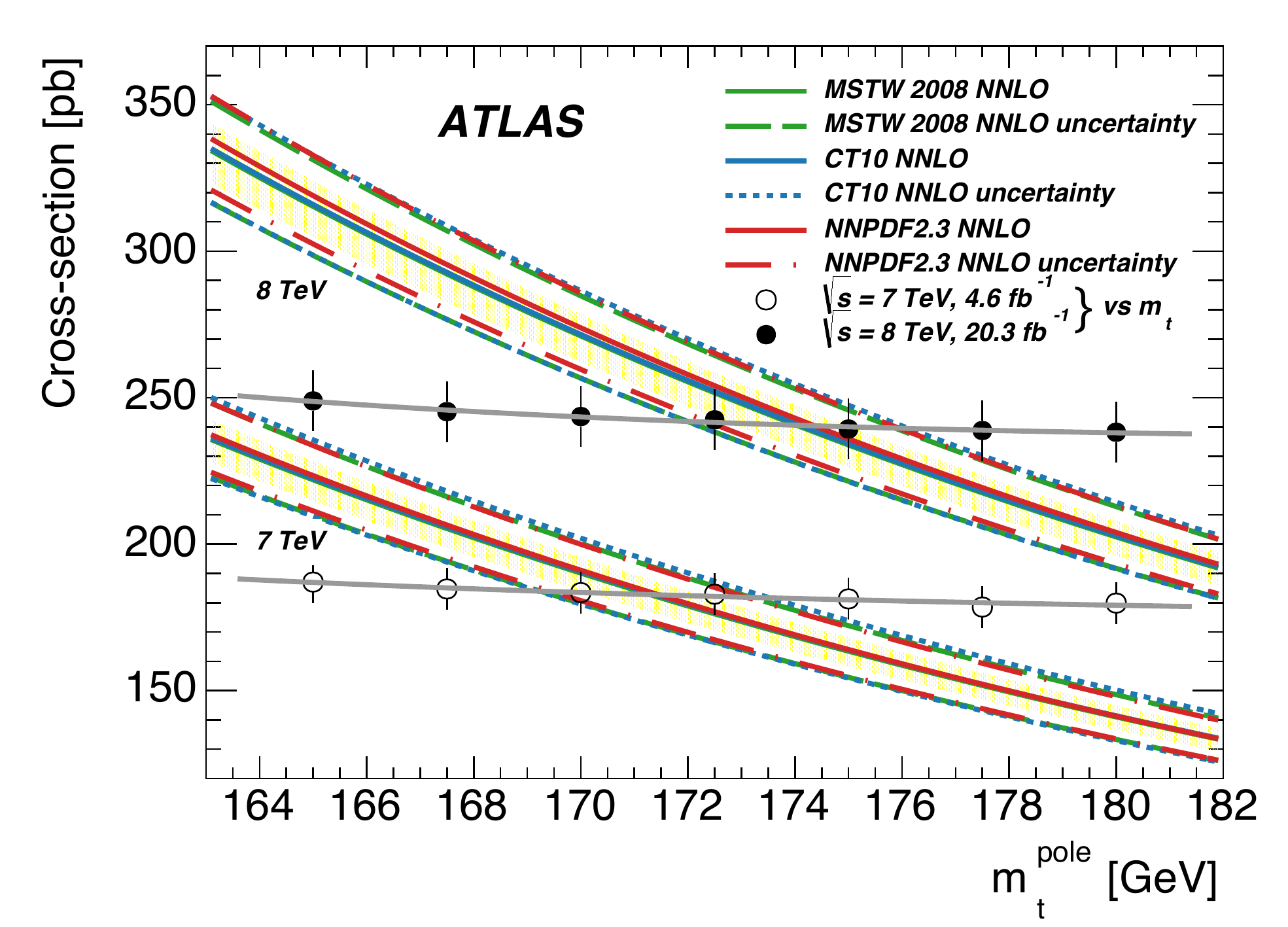}}
\end{minipage}
\caption[]{\textbf{Left}: Final distributions of the reconstructed $m_t$ variable from the single most precise CMS Run-1 $m_{top}$ measurement in the semileptonic \ttbar \ decay channel.\cite{cms_78tev_allchannels} \textbf{Right}: Theoretical dependence of the \ttbar \ production cross-section on the top-quark pole mass at both $\sqrt{s} = $7 and 8 \TeV, together with measured values from ATLAS.\cite{atlas_78tev_xs}}
\label{fig:mtop-plots}
\vspace{-0.3cm}
\end{figure}

Semileptonic \ttbar \ events offer an ideal balance between the ability to select candidate events \textit{via} the identification of a single high-\pT \ lepton ($\ell = e/\mu$) from a leptonic (anti-)top-quark decay, while fully reconstructing the hadronically decaying (anti-)top quark.  The $\rts=$8 \TeV \ analysis by CMS \cite{cms_78tev_allchannels} currently yields the single most precise LHC $m_{top}$ measurement, though tension exists between this and a recent D0 measurement.\cite{d0_semilep} The corresponding $m_{t}^{fit}$ distribution for the final selected events following all event selection cuts, together with the application of a goodness-of-fit weight for all permutations, is shown in the left plot of Figure~\ref{fig:mtop-plots}.  In addition, the analysis presents new measurements of $m_{top}$ as a function of several key variables.  A separate $\rts=$8 \TeV \ analysis focuses exclusively on the $\mu$+jets channel \cite{cms_8tev_semilep_best} and utilizes an $R$ observable similar to \rtt \ described in the all-hadronic ATLAS measurement.

The $\rts =$7 \TeV \ ATLAS measurement  in the semileptonic ($e/\mu$)\ channel \cite{atlas_7tev_semilep} employs a novel 3D template method in which additional observables are constructed, in addition to the reconstructed top-quark mass ($m_{top}^{reco}$), in order to reduce the contribution from the systematic uncertainties due to the jet and b-jet energy scales (JES and bJES, respectively).  Events are reconstructed using a kinematic fit together with the known \Wboson \ boson mass ($m_{\Wboson}$).

\vspace{-0.35cm}
\subsection{Charged hadrons in the dileptonic/semileptonic \ttbar \ channel}\label{sec:singletop}

A recent CMS measurement at $\rts = $8 \TeV \ identifies charged hadrons ($J/\psi$, $D^{0}$, and $D^{*}(2010)^{+}$) produced \textit{via} the decay of bottom-flavoured hadrons from the top quark decays. The reconstructed invariant mass of the lepton and secondary vertex system in signal events affords additional sensitivity to $m_{top}$ compared to the lepton information alone, while circumventing the larger systematic uncertainties associated with the JES and bJES.\cite{cms_8tev_sec_vertices}  A similar analysis, which is currently statistically limited, focuses on the identification of $J/\psi$ mesons alone \cite{cms_8tev_jpsi}.  In both analyses the systematic uncertainty associated with the $b$-quark fragmentation modelling becomes significant, while the jet-related systematic uncertainties are greatly diminished. 

\vspace{-0.35cm}
\subsection{Single-top t-channel}\label{sec:singletop}

Both the ATLAS and CMS collaborations have performed $m_{top}$ measurements at $\rts=$ 8 \TeV \ by employing a set of event selection cuts chosen specifically to target top quarks produced singly \textit{via} a t-channel electroweak interaction.\cite{atlas_8tev_singletop,cms_8tev_singletop} Such measurements provide complementarity to the set of standard analyses involving the strongly produced $t\bar{t}$ pairs.  Both the ATLAS ($e/\mu$ + jets) and CMS ($\mu$ + jets) analyses use a one-dimensional template method, where the choice of $m_{top}$-sensitive observable is the reconstructed mass of the lepton-$b$-quark system ($m_{\ell b}$) for ATLAS and the reconstructed invariant mass of the reconstructed top quark ($m_{\ell \nu b}$) for CMS.

\vspace{-0.3cm}
\section{Indirect measurements from production cross-sections}\label{sec:direct}

A second set of analyses involve measurements of the top-quark pole mass directly, a quantity whose value could differ from the $m_{top}$ value used in the generation of simulated Monte Carlo events by $\mathcal{O}$(1 \GeV).\cite{hoang_mtop}  In such analyses the theoretical uncertainties typically dominate.

\vspace{-0.35cm}
\subsection{Inclusive \ttbar \ production cross-section using dileptonic ($e\mu$) events}\label{sec:dilepxs}

Both ATLAS and CMS have performed precision measurements of the inclusive \ttbar \ production cross-section ($\sigma_{\ttbar}$) in dileptonic ($e\mu$) final states \cite{atlas_78tev_xs,cms_78tev_xs}.  The dependence of $\sigma_{\ttbar}$, based on theoretical (NNLO+NNLL) predictions, on the top-quark pole mass can be exploited in order to extract a measurement of $m_{top}$.  The right plot in Figure~\ref{fig:mtop-plots} shows an example of this dependence from the ATLAS analysis.  The largest sources of uncertainty are primarily associated with the choice of parton distribution functions (PDFs) for the proton, the values of the strong coupling constant ($\alpha_{s}$) and QCD factorization and renormalization scales, and the LHC beam energy.  

\vspace{-0.35cm}
\subsection{Differential \ttbar +1-jet production cross-section using semileptonic ($e/\mu$) events}\label{sec:ttj_xs}

A $\rts=$7 \TeV \ ATLAS analysis \cite{atlas_7tev_ttj_polemass} targets the theoretical sensitivity of the \ttbar \ + 1-jet production cross-section ($\sigma_{\ttbar+j}$) to the top-quark pole mass, specifically in the context of gluon radiation.  A fit is performed to the unfolded differential distribution of $\sigma_{\ttbar+j}$ with respect to a variable ($\rho_s$) associated with the invariant mass of the $\ttbar+$1-jet system, thereby returning an estimate of top-quark pole mass.  It is expected that performing a similar measurement at $\rts = 8$ \TeV \ would lead to improved precision, since at $\rts=$7 \TeV \ the statistical uncertainty is dominant.

\vspace{-0.4cm}
\section*{Conclusions and Outlook}

\begin{figure}
\vspace{-0.5cm}
\begin{minipage}[c]{0.45\linewidth}
\centerline{\includegraphics[width=1.\linewidth]{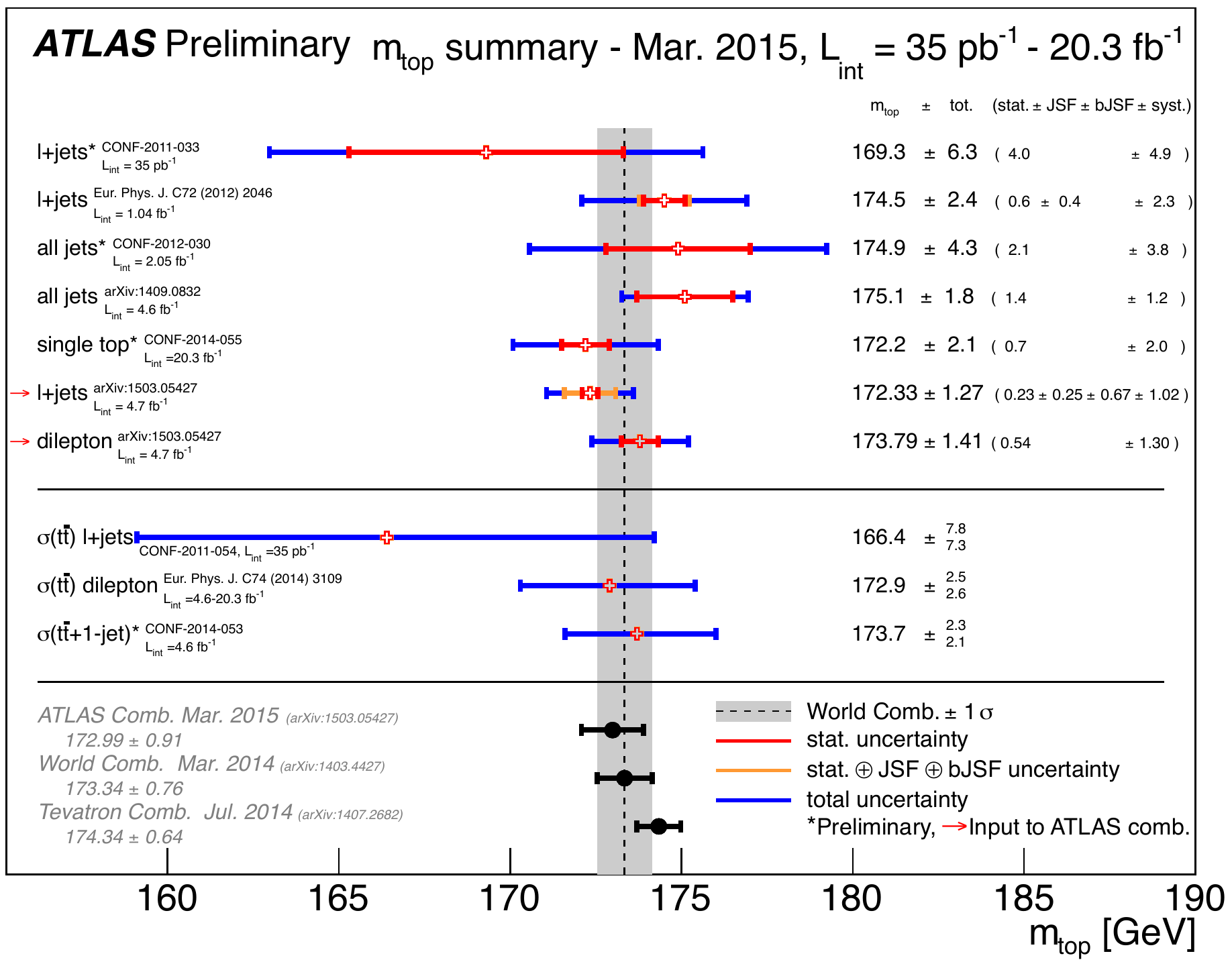}}
\end{minipage}
\hfill
\begin{minipage}[c]{0.45\linewidth}
\hspace{0.2cm}
\centerline{\includegraphics[width=0.68\linewidth]{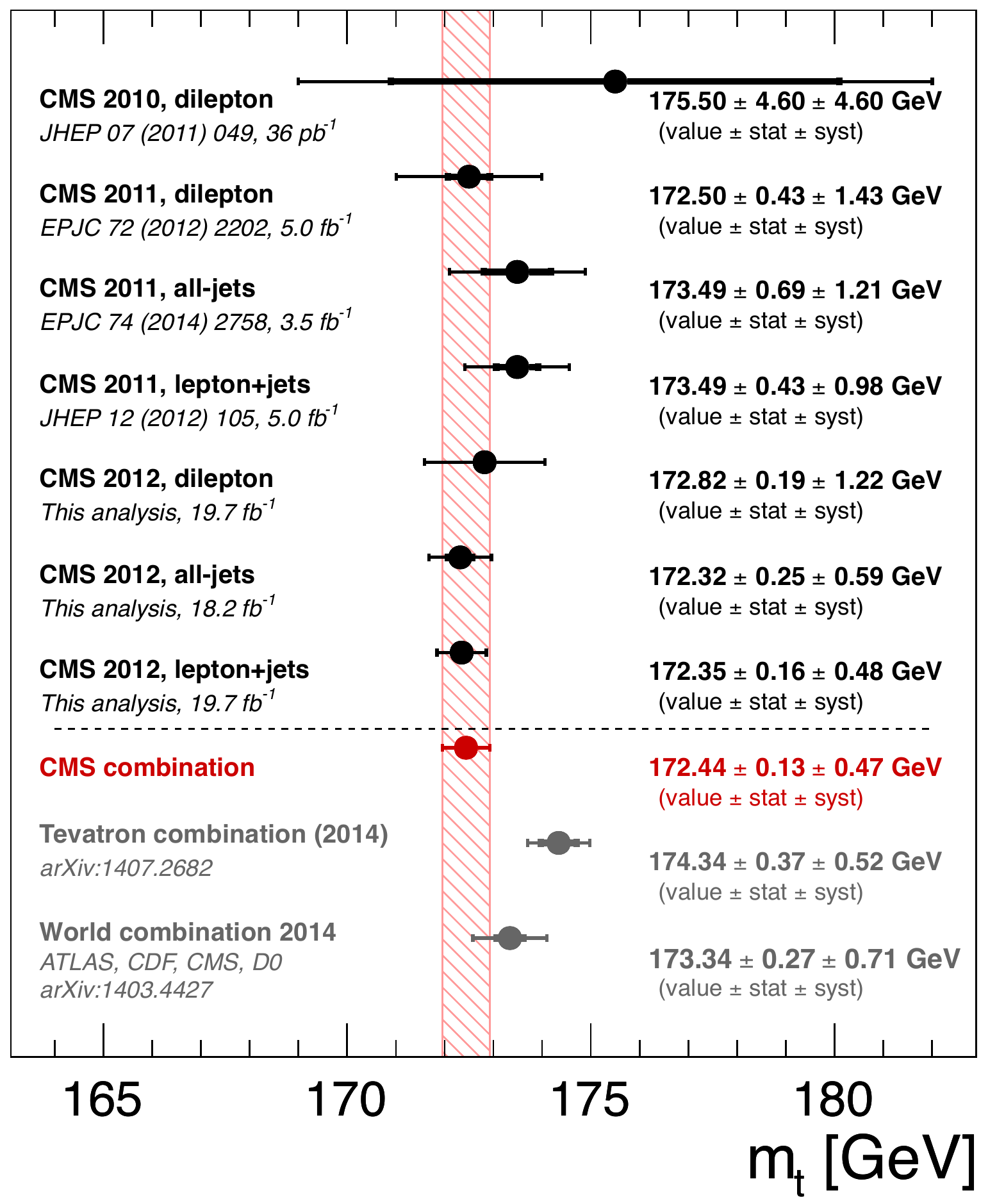}}
\end{minipage}
\caption[]{Summary plots for the standard ATLAS (\textbf{left}) and CMS (\textbf{right}) $m_{top}$ measurements.\cite{atlas_public,cms_78tev_allchannels}  The ATLAS plot includes the $\rts=7$ \TeV \ measurements performed in the standard $\ttbar$ decay channels as well as the indirect measurements from the cross-sections at both $\rts=7$ and 8 \TeV.  The CMS plot includes the results from the standard \ttbar \ decay channels at both $\rts=7$ and 8 \TeV.  The measurements performed using single-top events and charged hadrons, as well as the indirect measurements from the cross-section, are not included in the CMS plot.}
\label{fig:combinations}
\vspace{-0.5cm}
\end{figure}

This overview highlights the breadth of approaches and analysis strategies employed by both the ATLAS and CMS collaborations during Run 1 at the LHC in order to perform high-precision $m_{top}$ measurements.  The most recent summary plots from both experiments at the time of writing are shown in Figure~\ref{fig:combinations}.  Both experiment-specific and LHC combinations of results have been performed.\cite{cms_run1_comb,lhc_7tev_comb}  The combined ATLAS result from the $\rts=7$ \TeV \ dileptonic and semileptonic \ttbar \ measurements yields a value of $m_{top} = 172.99 \pm 0.48$ (stat) $\pm 0.78$ (syst) \GeV. Additional precision is to be expected with the inclusion of $\rts=$ 8 \TeV \ measurements.  The combined CMS result of $m_{top} = 172.44 \pm 0.13$ (stat) $\pm 0.47$ (syst) \GeV \ is based on all $\rts=$7 and 8 \TeV \ measurements performed in the three standard \ttbar \ decay channels.  The results of both experiments are in good agreement.  

While future measurements at $\rts=13$ \TeV \ will recognizably benefit from higher statistics, the reduction of several theoretical uncertainties would further allow for improved precision in both the indirect measurements from the production cross-sections and those analyses which avoid employing jets in the construction of an observable sensitive to $m_{top}$.

\end{document}